\documentstyle[11pt,epsfig]{article}
\newcounter{subequation}[equation]

\makeatletter

\def\thesubequation{\theequation\@alph\c@subequation}
\def\@subeqnnum{{\rm (\thesubequation)}}
\def\slabel#1{\@bsphack\if@filesw {\let\thepage\relax
   \xdef\@gtempa{\write\@auxout{\string
      \newlabel{#1}{{\thesubequation}{\thepage}}}}}\@gtempa
   \if@nobreak \ifvmode\nobreak\fi\fi\fi\@esphack}
\def\subeqnarray{\stepcounter{equation}
\let\@currentlabel=\theequation\global\c@subequation\@ne
\global\@eqnswtrue
\global\@eqcnt\z@\tabskip\@centering\let\\=\@subeqncr
$$\halign to \displaywidth\bgroup\@eqnsel\hskip\@centering
  $\displaystyle\tabskip\z@{##}$&\global\@eqcnt\@ne
  \hskip 2\arraycolsep \hfil${##}$\hfil
  &\global\@eqcnt\tw@ \hskip 2\arraycolsep
  $\displaystyle\tabskip\z@{##}$\hfil
   \tabskip\@centering&\llap{##}\tabskip\z@\cr}
\def\endsubeqnarray{\@@subeqncr\egroup
                     $$\global\@ignoretrue}
\def\@subeqncr{{\ifnum0=`}\fi\@ifstar{\global\@eqpen\@M
    \@ysubeqncr}{\global\@eqpen\interdisplaylinepenalty \@ysubeqncr}}
\def\@ysubeqncr{\@ifnextchar [{\@xsubeqncr}{\@xsubeqncr[\z@]}}
\def\@xsubeqncr[#1]{\ifnum0=`{\fi}\@@subeqncr
   \noalign{\penalty\@eqpen\vskip\jot\vskip #1\relax}}
\def\@@subeqncr{\let\@tempa\relax
    \ifcase\@eqcnt \def\@tempa{& & &}\or \def\@tempa{& &}
      \else \def\@tempa{&}\fi
     \@tempa \if@eqnsw\@subeqnnum\refstepcounter{subequation}\fi
     \global\@eqnswtrue\global\@eqcnt\z@\cr}
\let\@ssubeqncr=\@subeqncr
\@namedef{subeqnarray*}{\def\@subeqncr{\nonumber\@ssubeqncr}\subeqnarray}
\@namedef{endsubeqnarray*}{\global\advance\c@equation\m@ne%
                           \nonumber\endsubeqnarray}

\def\dalemb#1#2{{\vbox{\hrule height .#2pt
        \hbox{\vrule width.#2pt height#1pt \kern#1pt
                \vrule width.#2pt}
        \hrule height.#2pt}}}
\def\square{\mathord{\dalemb{6.8}{7}\hbox{\hskip1pt}}}

    \let\e=\epsilon
  \let\q=\theta  
  \let\n=\nu

\def\nn{\nonumber} \def\bd{\begin{document}} \def\ed{\end{document}}
\def\ds{\documentstyle} \let\fr=\frac \let\bl=\bigl \let\br=\bigr
\let\Br=\Bigr \let\Bl=\Bigl
\let\bm=\bibitem
\let\na=\nabla
\let\pa=\partial \let\ov=\overline
\def\ie{{\it i.e.\ }}
\newcommand{\be}{\begin{equation}}
\newcommand{\ee}{\end{equation}}
\def\ba{\begin{array}}
\def\ea{\end{array}}
\def\ft#1#2{{\textstyle{{\scriptstyle #1}\over {\scriptstyle #2}}}}
\def\fft#1#2{{#1 \over #2}}
\def\del{\partial}
\def\sst#1{{\scriptscriptstyle #1}}
\def\oneone{\rlap 1\mkern4mu{\rm l}}
\def\e7{E_{7(+7)}}
\def\td{\tilde}
\def\wtd{\widetilde}
\def\im{{\rm i}}
\def\bog{Bogomol'nyi\ }
\def\q{{\tilde q}}
\def\hast{{\hat\ast}}
\def\0{{\sst{(0)}}}
\def\1{{\sst{(1)}}}
\def\2{{\sst{(2)}}}
\def\3{{\sst{(3)}}}
\def\4{{\sst{(4)}}}
\def\5{{\sst{(5)}}}
\def\6{{\sst{(6)}}}
\def\7{{\sst{(7)}}}
\def\8{{\sst{(8)}}}
\def\n{{\sst{(n)}}}
\def\oo{{\"o}}
\def\hA{\hat{\cal A}}
\def\ns{{\sst {\rm NS}}}
\def\rr{{\sst {\rm RR}}}
\def\tH{{\widetilde H}}
\def\tB{{\widetilde B}}
\def\cA{{\cal A}}
\def\cF{{\cal F}}
\def\tF{{\wtd F}}
\def\Z{\rlap{\sf Z}\mkern3mu{\sf Z}}
\def\ep{{\epsilon}}
\def\IIA{{\rm IIA}}
\def\IIB{{\rm IIB}}
\def\ads{{\rm AdS}}
\def\R{\rlap{\rm I}\mkern3mu{\rm R}}
\def\mapright#1{\smash{\mathop{-\!\!\!-\!\!\!-\!\!\!-\!\!\!-\!\!\!
             \longrightarrow}\limits^{#1}}}

\def\Ei{{\hbox{Ei}}}
\def\Ci{{\hbox{Ci}}}
\def\Si{{\hbox{Si}}}

\newcommand{\ho}[1]{$\, ^{#1}$}
\newcommand{\hoch}[1]{$\, ^{#1}$}
\newcommand{\bea}{\begin{eqnarray}}
\newcommand{\eea}{\end{eqnarray}}
\newcommand{\ra}{\rightarrow}
\newcommand{\lra}{\longrightarrow}
\newcommand{\Lra}{\Leftrightarrow}
\newcommand{\aap}{\alpha^\prime}
\newcommand{\bp}{\tilde \beta^\prime}
\newcommand{\tr}{{\rm tr} }
\newcommand{\Tr}{{\rm Tr} }
\newcommand{\NP}{Nucl. Phys. }
\newcommand{\tamphys}{\it Center for Theoretical Physics,
Texas A\&M University, College Station, TX 77843}

\newcommand{\upenn}{\it Department of Physics and Astronomy,\\ University
of Pennsylvania, Philadelphia, PA 19104}

\newcommand{\brussels}{\it Physique Th\'eorique et Math\'ematique,
Universit\'e Libre de Bruxelles,\\ Campus Plaine C.P. 231, B-1050
Bruxelles, Belgium}

\newcommand{\auth}{H. L\"u\hoch{\dagger1} and J.F.
V\'azquez-Poritz\hoch{\ddagger2}}

\begin{document}
\begin{flushright}
MCTP-02-12\\
ULB-TH/02-05\\
February  2002\\
\hfill{\bf hep-th/0202175}\\
\end{flushright}

\vspace{10pt}

\begin{center}

{\large {\bf $S^1$-wrapped D3-branes on Conifolds}}

\vspace{20pt}
\auth

\vspace{10pt}
{\hoch{\dagger}\it Michigan Center for Theoretical Physics\\
University of Michigan, Ann Arbor, Michigan 48109}

\vspace{10pt}
{\hoch{\ddagger}\brussels}

\vspace{30pt}

\underline{ABSTRACT}
\end{center}

          We construct a D3-brane wrapped on $S^1$, which is fibred
over the resolved conifold as its transverse space.  Whereas a
fractional D3-brane on the resolved conifold is not supersymmetric and
has a naked singularity, our solution is supersymmetric and regular
everywhere.  We also consider an $S^1$-wrapped D3-brane on the
resolved cone over $T^{1,1}/Z_2$, as well as on the deformed conifold.
In the former case, we obtain a regular supergravity dual to a certain
four-dimensional field theory whose Lorentz and conformal symmetries
are broken in the IR region and restored in the UV limit.

{\vfill\leftline{}\vfill
\vskip 10pt
\footnoterule {\footnotesize \hoch{1}
Research supported in full by DOE grant DE-FG02-95ER40899.

{\footnotesize \hoch{2} Research supported in full by the Francqui
Foundation (Belgium), the Actions de Recherche Concert{\'e}es
\phantom{of the} of the
Direction de la Recherche Scientifique - Communaut\'e Francaise de
Belgique, IISN-Belgium
\phantom{of the} (convention 4.4505.86).  }
\vskip  -12pt}

\pagebreak
\setcounter{page}{1}


\section{Introduction}

      D3-branes no doubt provide the most natural framework for the
study of strongly-coupled four-dimensional Yang-Mills theory from the
points of view of supergravity and string theory, {\it via} the
AdS/CFT correspondence \cite{malda,gkp,wit}.  In order to reduce the
supersymmetry to a minimum, one can replace the six-dimensional
Euclidean transverse space with a Calabi-Yau manifold.\footnote{An
analogous construction was proposed earlier in \cite{dlps} for the
M2-brane.}  The simplest example of six-dimensional Calabi-Yau
manifolds is the (non-compact and singular) conifold, defined as a
cone over $T^{1,1}=(S^3\times S^3)/S^1$. The near-horizon geometry of
the D3-brane now becomes AdS$_5\times T^{1,1}$, which provides a
supergravity dual to the ${\cal N}=1$, $D=4$ superconformal Yang-Mills
theory. There is a supersymmetric 2-cycle in the $T^{1,1}$ space, upon
which one can wrap additional D5-branes or NS5-branes, giving rise to
supersymmetric fractional D3-branes
\cite{klebtsey,klebstra,ganpol,gubser}.  The conformal symmetry is
broken by the distance-dependent logarithmic contribution to the
D3-brane charge.  The solution, however, has a short-distance naked
singularity and hence provides a structural behavior only at large
distance, corresponding to the UV (ultra-violet) region of the dual
field theory.

        In the construction of the fractional D3-brane of type IIB
theory, the complex 3-form $F_\3=F_\3^{\rm RR} + {\rm i}\, F_\3^{\rm
NS}$ is set proportional to the complex self-dual 3-form $\omega_\3$
of the conifold, such that it contributes non-trivially to the Bianchi
identity of the 5-form: $dF_\5=\ft12{\rm i}\, \bar F_\3\wedge F_\3$.
To resolve the above naked singularity, $\omega_\3$ must be square
integrable at short distance. This requires that the conifold must
have a non-collapsing 3-cycle \cite{cglpns2d2}. There are two
smoothed-out versions of the conifold, namely the deformed conifold
and the resolved conifold \cite{candel}.  In the former case, the
singular apex is blown up to a smooth three-sphere, and hence it has a
non-collapsing 3-cycle.  The fractional D3-brane on the deformed
conifold was constructed in \cite{klebstra}. This solution is
supersymmetric and regular everywhere.  On the other hand, in the case
of the resolved conifold, the singular apex is blown up to a smooth
two-sphere. Thus, it has a non-collapsing 2-cycle, but a collapsing
3-cycle. The fractional D3-brane over the resolved conifold was
constructed in \cite{zaytse}, and was shown to have a repulson-like
naked singularity.  Furthermore, it was shown to be non-supersymmetric
\cite{cglptrans,cglpsten}.

       Since the resolved conifold has a non-collapsing 2-cycle, the
solution would be regular if it was the harmonic 2-form instead of the
3-form to provide the D3-brane source.  Following the technique
developed in \cite{lvres}, we consider the D3-brane wrapped on $S^1$,
which is fibred over the resolved conifold.  Besides the K\"ahler
form, there are two additional harmonic 2-forms supported by the
resolved conifold.  One of them is square integrable at short distance
and the resulting solution is regular everywhere.  The harmonic 2-form
carries non-trivial Taub-NUT type flux.  Consequently, it is not
normalizable at large distance, and contributes a distance-dependent
D3-brane charge.  The other 2-form falls off rapidly at infinity and
does not carry any flux, which implies that the D3-brane charge is
well-defined. However, it is not square integrable at short distance
and hence the solution is singular in that region.  We show that both
solutions are supersymmetric, preserving the minimal amount of
supersymmetry.

        Recently, it was shown that the cone over $T^{1,1}/Z_2$ can
also be resolved.  The metric was obtained in \cite{b2,p2,pando2}.  It
describes a complex-line bundle over $S^2\times S^2$.  In this case,
the singular apex is blown up to a smooth $S^2\times S^2$, and hence
the manifold has a non-collapsing 4-cycle.  On the other hand, there
are no 4-cycles in $T^{1,1}/Z_2$. It follows that the 4-form does not
carry non-trivial charge, and hence is fully normalizable. In six
dimensions, a 4-form is Hodge dual to a 2-form.  This enables us to
construct a regular $S^1$-wrapped D3-brane on the resolved cone over
$T^{1,1}/Z_2$, which we show to be supersymmetric.  In this
configuration, the conformal and Lorentz symmetries of the original,
unwrapped D3-brane are broken, but both are restored at large
distance.

        This paper is organized as follows. In section 2, we discuss
the general construction of a D3-brane wrapped on $S^1$, which is
fibred over the six-dimensional transverse space. By T-dualizing and
lifting the solution to eleven dimensions, we find the condition for
which the supersymmetry is preserved.  We show that the $S^1$-wrapped
D3-brane and fractional D3-brane have a common origin as the modified
supermembrane in M-theory.  In section 3, we consider the case in
which the transverse space is a conifold.  For this, a fractional
D3-brane is singular, whereas an $S^1$-wrapped D3-brane is regular
everywhere.  In sections 4 and 5, we find regular $S^1$-wrapped
D3-brane solutions on the resolved conifold over $T^{1,1}$ or
$T^{1,1}/Z_2$, respectively.  The latter solution is of particular
interest since both conformal and Lorentz symmetries are broken in the
IR region of the dual field theory, and restored in the UV limit.  On
the other hand, a D3-brane wrapped over $S^1$ on the deformed conifold
has a naked singularity at short distance, as we see in section 6. We
present conclusions in section 7.

\section{$S^1$ wrapped D3-brane}

        The D3-brane of type IIB supergravity is supported by the
self-dual 5-form field strength, with a six-dimensional Ricci-flat
transverse space. Due to the Bianchi identity $dF_\5=F_\3^{\rm
NS}\wedge F_\3^{\rm RR}$, one can construct a fractional D3-brane if
the transverse space has a self-dual 3-cycle.  If instead the
transverse space has a 2-cycle $L_\2$, we can construct an
$S^1$-wrapped D3-brane with one of the world-volume coordinates fibred
over the transverse space.  Using the same technique developed in
\cite{lvres}, we find that the solution is given by
\bea
ds_{10}^2 &=& H^{-\ft12}\, \Big(-dt^2 + dx_1^2 + dx_2^2 +
(dx_3 +\cA_\1)^2\Big) + H^{\ft12}\, ds_6^2\,,\nn\\
F_\5&=&dt\wedge dx_1\wedge dx_2\wedge (dx_3+\cA_\1)\wedge dH^{-1} -
{\ast_6 dH}\nn\\
&&+ m\, {\ast_6 L_\2}\wedge (dx_3+\cA_\1) +
dt\wedge dx_1\wedge dx_2\wedge L_\2\,,\nn\\
d\cA_\1&=&m\, L_\2\,,\label{wrapd3gen}
\eea
where $L_\2$ is a harmonic 2-form in the transverse space of the
metric $ds_6^2$, and $\ast_6$ is the Hodge dual with respect to
$ds_6^2$.  The equations of motion are satisfied, provided that
\be
\square H = -\ft12 m^2 L_\2^2\,,\label{lapmod}
\ee
where $\square$ is the Laplacian in $ds_6^2$.

        A convenient method of determining the preserved supersymmetry
of the solution is to T-dualize the fibre coordinate $x_3$ to obtain a
modified D2-brane in type IIA theory and then dimensionally oxidize
the solution to $D=11$.  The modified D2-brane is given by
\bea
ds_{10}^2 &=& H^{-\ft58}\, (-dt^2 + dx_1^2 + dx_2^2) + H^{\ft38}\,
(ds_6^2 + dz_1^2)\,,\nn\\
F_\4 &=& dt\wedge dx_1\wedge dx_2\wedge dH^{-1} + {\ast_6
L_\2}\,,\quad F_\3 = m\,L_\2\wedge dz_1\,,\quad
\phi = \ft14 \log(H)\,.
\eea
Note that, under T-duality, the fibre coordinate is untwisted and
corresponds to the $z_1$ coordinate of the transverse space.  Lifting
to $D=11$ yields the modified M2-brane, which is given by
\bea
ds_{11}^2 &=& H^{-\ft23}\, (-dt^2 + dx_1^2 + dx_2^2) + H^{\ft13}
\, (ds_6^2 + dz_1^2 + dz_2^2)\,,\nn\\
F_\4 &=& dt\wedge dx_1\wedge dx_2\wedge dH^{-1} + m\, L_\4\,,
\eea
where
\be L_\4 = {\ast_6 L_\2} + L_\2\wedge dz_1\wedge dz_2 \ee
is a self-dual harmonic 4-form living in the 8-dimensional Ricci-flat
transverse space with the metric $ds_6^2 + dz_1^2 + dz_2^2$.  This
type of modification to the M2-brane, which makes use of the
interaction in $d{\ast F_\4}=\ft12 F_\4 \wedge F_\4$, has been
considered in \cite{deklm,htr,becker,cglptrans} (see also, {\it e.g.,}
\cite{beckers,cglpsten,hk,cglpns2d2,cglphyper,cglpspin7,cgllp}).  The
introduction of $L_\4$ to the M2-brane solution preserves all of the
initial supersymmetries, provided that \cite{becker}
\be
L_{abcd}\, \Gamma^{bcd}\, \epsilon=0\,,\label{susycon}
\ee
where $\epsilon$ is a Killing spinor in the transverse space.

     Applying the same procedure to the fractional D3-brane yields a
modified M2-brane with
\be
L_\4= G_\3\wedge dz + \bar L_\3 \wedge d{\bar z}\,.
\ee
where $G_\3$ is a complex self-dual 3-form in $ds_6^2$ and $z=z_1 +
{\rm i}\, z_2$.  Thus, the wrapped D3-brane and fractional D3-brane
can be united in a Spin(7) manifold with the metric $ds_6^2 + dz_1^2 +
dz_2^2$ as its Gromov-Hausdorff limit.

\section{On the conifold}

         The simplest Calabi-Yau manifold in $D=6$ is the conifold
$ds_6^2 = dr^2 + r^2\, ds_{T^{1,1}}^2$, which is a Ricci-Flat cone
over the Einstein space $T^{1,1}$
\bea
ds_{T^{1,1}}^2 =\lambda^2\, (d\psi + \cos\theta\, d\phi -
\cos\td\theta\,d\td\phi)^2 + 
\ft16 (d\theta^2 + \sin^2\theta\, d\phi^2) +
\ft16 (d\td\theta^2 + \sin^2\td\theta\, d\td\phi^2)\,.\label{t11}
\eea
The constant $\lambda$ measures the squashing of the $U(1)$ fibre
coordinate $\psi$.  For the $T^{1,1}$ space to be Einstein, one must
have $\lambda=1/3$.  The period of $\psi$ is $4\pi$.  If instead the
period is $4\pi/n$, the space is $T^{1,1}/Z_n$.  In the $T^{1,1}$,
there is a supersymmetric 2-cycle $\omega_\2$ and its dual 3-cycle
$\omega_\3$, which are given by
\be
\omega_\2 = \Omega_\2 + \wtd\Omega_\2\,,\qquad
\omega_\3 = \ft13 (d\psi +\cos\theta\, d\phi
-\cos\td\theta\,d\td\phi) \wedge (\Omega_\2 + \wtd \Omega_\2 )\,,
\ee
where $\Omega_\2$ and $\wtd \Omega_\2$ are the volume-forms of the two
$S^2$.  Thus, the conifold supports a harmonic 2-form and complex
self-dual 3-form
\be
L_\2 = \ft16\omega_\2\,,\qquad
G_\3 = \omega_\3 + {\rm i}\, \omega_\2\wedge \fft{dr}{r}\,.
\label{23forms}
\ee
There are two ways of modifying a D3-brane on a conifold.  The first
is to equate the complex 3-form field strength in type IIB
supergravity to the above self-dual 3-form \cite{klebtsey}.  Since
${\cal R}e(G_\3)=\omega_\3$ carries non-trivial flux, this describes
an additional D5-brane wrapped on the supersymmetric 2-cycle, and
hence it is called a fractional D3-brane.  The modification to the
harmonic function $H$ is a logarithmic contribution from the 5-brane
charge:
\be H= 1+ \fft{Q+ 15 m^2\log r}{r^4}\,.\label{fracd3} \ee
This case has been extensively studied. The solution has a naked
singularity at small distance and resolutions have been proposed.  As
discussed in the introduction, there are two resolutions to the
conifold itself, namely the deformed conifold and the resolved
conifold \cite{candel}.  In order to have regular small-distance
behavior in $H$, it is clear that the 3-cycle should be
non-collapsing.  This is the case for the deformed conifold but not
for the resolved conifold.  Thus, there is no regular fractional
D3-brane on the resolved conifold constructed so far \cite{zaytse}. In
fact, the solution is non-supersymmetric \cite{cglptrans,cglpsten}.

      In this paper, we instead consider a wrapped D3-brane on $S^1$
which is fibred over the conifold, as described in (\ref{wrapd3gen}).
In this case, the function $H$ is modified in power law of $r$:
\be H=1+ \fft{m^2}{4r^2} + \fft{Q}{r^4}\,.\label{nutd3} \ee
The solution is regular everywhere already.  Even though the $1/r^4$
term dominates at small $r$, the conformal symmetry is broken
everywhere due to the fibration which carries non-vanishing charge
\be
\int_{r\rightarrow\infty} L_\2 = Q_2 \ne 0\,.
\ee
This charge is analogous to the Taub-NUT charge. 

        In addition to the $L_\2$ given in (\ref{23forms}), the
conifold supports another harmonic 2-form, given by
\be
\wtd L_\2 = \fft{2}{3r^5}\, dr\wedge (d\psi +\cos\theta\, d\phi
-\cos\td\theta\,d\td\phi) +
\fft{1}{6r^4} (\Omega_\2 -\wtd \Omega_\2)\,,\label{2nd2form}
\ee
Now the function $H$ is modified to
\be H=1+ \fft{Q}{r^4} -\fft{\wtd m^2}{20r^{10}}\,.\label{nonutd3}
\ee
This is very different from the above $S^1$-wrapped D3-brane.  The
fibration does not carry any charge, namely
\be
\int_{r\rightarrow\infty} \wtd L_\2 = 0.
\ee
In fact, the fibration rapidly vanishes at large $r$.  As a
consequence, the solution becomes AdS$_5\times T^{1,1}$ at large $r$,
and hence both the conformal and Lorentz symmetries are restored.  It
has a naked singularity at small but finite $r=r_0$.  As we shall see
in section 5, when the principal orbit is replaced with $T^{1,1}/Z_2$,
this naked singularity can be resolved.

        Finally there is the K\"ahler form.  Its contribution to the
function $H$ is so badly behaved at large distance that there is an
unresolvable naked singularity.  Furthermore, as we shall see in
section 4, it will break the supersymmetry.  We shall not consider the
K\"ahler form in this paper.

\section{On the resolved conifold}

      The metric of the resolved conifold over $T^{1,1}$ is given by
\bea ds_6^2 = d\rho^2 + a^2\, (\Sigma_1^2 + \Sigma_2^2) + b^2\,
(\sigma_1^2 + \sigma_2^2) + c^2\, (\Sigma_3-\sigma_3)^2.
\label{rescon} \eea
The functions $a$, $b$ and $c$ depend only on the radial variable
$\rho$; $\sigma_i$ and $\Sigma_i$ are left-invariant 1-forms of
$SU(2)\times SU(2)$.  They can be expressed in terms of Euler angles
as
\bea
&&\sigma_1+\im\, \sigma_2= e^{-\im\, \psi}\, (d\theta + \im\,
\sin\theta\, d\phi)\,,\qquad \sigma_3=d\psi + \cos\theta\,
d\phi\,,\nn\\
&&\Sigma_1+\im\, \Sigma_2= e^{-\im\, \wtd\psi}\, (d\td\theta + \im\,
\sin\td\theta\, d\td\phi)\,,\qquad \Sigma_3=d\wtd\psi + \cos\td\theta\,
d\td\phi\,,
\eea
and they satisfy
\be
d\sigma_i=-\ft12 \ep_{ijk}\, \sigma_j\wedge \sigma_k\,,\qquad
d\Sigma_i=-\ft12 \ep_{ijk}\, \Sigma_j\wedge \Sigma_k\,.
\ee
Note that, although there are ostensibly six coordinates here,
when one substitutes them into (\ref{rescon}), $\psi$ and
$\wtd\psi$ appear only through the combination $\psi-\wtd\psi$.

       The existence of Killing spinors implies that functions $a$,
$b$ and $c$ satisfy the following first-order equations:
\bea
2a\,\dot a = c = 2b\,\dot b\,,\qquad
\dot c= 1- \fft{c^2}{2a^2} - \fft{c^2}{2b^2}\,.
\label{resolvefo}
\eea
Here, a dot denotes a derivative with respect to $\rho$. The
solution for the resolved conifold is given by
\bea
&&a^2=\ft16 r^2\,,\qquad
b^2=\ft16 (r^2 + 6\ell^2)\,,\qquad c^2=\fft{r^2}{9h^2}\,,\nn\\
&& h^2=\fft{r^2 + 6\ell^2}{r^2 + 9\ell^2}\,,\qquad d\rho=h\, dr,
\eea
where we have introduced a more convenient radial variable $r$, which
runs from 0 to $\infty$.  As $r$ approaches 0, the metric becomes
$R^4\times S^2$, which is the topology of the manifold.
Asymptotically at large distance, the metric becomes a cone over
$T^{1,1}$.

     We are interested in finding a harmonic 2-form supported by this
metric. The most general ansatz for a 2-form with respect to the
isometry of (\ref{rescon}) is given by
\be
L_\2 = u_1\, e^0\wedge e^3 + u_2\, e^1\wedge e^2 + u_3\, e^4\wedge e^5
\,,\label{res2formans}
\ee
expressed in the vielbein basis $e^0=h\, dr$, $e^1=a\, \Sigma_1$,
$e^2=a\, \Sigma_2$, $e^3=c\, (\Sigma_3-\sigma_3)$, $e^4=b\, \sigma_1$
and $e^5=b\, \sigma_2$.  The closure and co-closure of $L_\2$ yield
the following solution:
\bea
u_1&=&-c_3 -\fft{6\ell^2\, c_1}{(r^2 + 6\ell^2)^2} +
\fft{2(r^2 +3\ell^2)\, c_2}{r^4\,(r^2+6\ell^2)}\,,\nn\\
u_2&=& c_3 + \fft{c_1}{r^2 + 6\ell^2} + \fft{c_2}{r^4\, (r^2 +
6\ell^2)}\,,\nn\\
u_3&=& -c_3 + \fft{(r^2 + 12\ell^2)\, c_1}{(r^2 + 6\ell^2)^2} -
\fft{c_2}{r^2\, (r^2 + 6\ell^2)^2}\,.\label{u}
\eea
Clearly, the harmonic 2-form associated with the $c_3$ terms is the
K\"ahler form.  The one associated with the $c_1$ terms, given by
\be L_\2 = -\fft{6\ell^2}{(r^2+6\ell^2)^2}\, e^0\wedge e^3 +
\fft{1}{r^2 + 6\ell^2}\, e^1\wedge e^2 + \fft{r^2 + 12\ell^2}{(r^2
+6\ell^2)^2}\, e^4\wedge e^5, \ee
gives rise to the $L_\2$ in (\ref{23forms}) at large $r$.  Hence
it has a non-trivial flux.  The square of this form is
\be
L_\2^2=\fft{4(r^4+18\ell^2\, r^2 + 108\ell^4)}{(r^2 + 6\ell^2)^4}\,.
\ee
Thus, it is square integrable for $r\rightarrow 0$ but not
normalizable at large distance.  There exists a regular solution to
$H$ in (\ref{lapmod}) which, after choosing appropriate integration
constants, is given by
\be
H=1 + \fft{m^2}{4(r^2 + 6\ell^2)}\,.
\ee
The solution is regular everywhere.  At small distance $r\rightarrow
0$, the function $H$ is a constant and at large distance, $H$ behaves
like (\ref{nutd3}).

      The 2-form associated with the $c_2$ terms in (\ref{u}) is given
by
\be
\wtd L_\2 = \fft{2(r^2 + 3\ell^2)}{r^4\, (r^2 + 6\ell^2)^2}\,
e^0\wedge e^3+\fft{1}{r^4\, (r^2 + 6\ell^2)}\, e^1\wedge e^2 -
\fft{1}{r^2\,(r^2+6\ell^2)^2}\, e^4\wedge e^5\,.
\ee
This gives rise to (\ref{2nd2form}) at large $r$.  We find that
\be
\wtd L_\2^2 = \fft{12(r^4 + 6\ell^2\, r^2 + 12\ell^4)}{r^8\,
(r^2 + 6\ell^2)^4}\,,
\ee
which falls off rapidly at large distance but is not square integrable
at small distance.  The resulting wrapped D3-brane solution has a
naked singularity.

      Both of the wrapped D3-branes are supersymmetric.  To see this
we first note that, after setting $c_3=0$, the three $u_i$'s
satisfy the following linear relation:
\be
u_1 - u_2 + u_3 =0\,.\label{u123res}
\ee
This linear dependence is crucial for the preservation of
supersymmetry.  To demonstrate this, we perform T-duality on the
wrapped coordinate and lift the solution to $D=11$, as discussed in
section 2.  It is straightforward to verify that the supersymmetric
condition (\ref{susycon}) precisely implies (\ref{u123res}).  It is
instructive to examine the case of the fractional D3-brane, for which
the complex self-dual harmonic 3-form is given by $G_\3=L\3 + {\rm i}
{\ast_6L_\3}$, where
\be
L_\3 = \fft{1}{c\, a^2}\, e^3\wedge e^1\wedge e^2 +
\fft{1}{c\, b^2}\, e^3\wedge e^4\wedge e^5\,.
\ee
In this case, the vielbein components of $G_\3$ are not linearly
dependent, except at $r=\infty$.  Thus, as shown in
\cite{cglptrans,cglpsten}, this self-dual 3-form cannot satisfy the
supersymmetric condition
\be
L_{abc}\, \Gamma^{abc}\, \epsilon=0\,,\qquad 
({\ast_6 L})_{abc}\, \Gamma^{abc}\,\epsilon=0,
\ee
obtained in \cite{gubser}.

        It should be emphasized that a linear dependency of the
vielbein components of a harmonic form is only a necessary condition,
but not sufficient.  Clearly, the vielbein components of the K\"ahler
form are linearly dependent since they are constants; however, the
resulting solution is not supersymmetric since the specific
relationship (\ref{susycon}) is not satisfied.

       In this section we have found that, due to the existence of the
non-collapsing 2-cycle in the resolved conifold, there exists a
square-integrable harmonic 2-form at short distance.  This yields a
regular and supersymmetric $S^1$-wrapped D3-brane.

\section{On the resolved cone over $T^{1,1}/Z_2$}

     The first-order equations (\ref{resolvefo}) admit a more general
regular solution:
\bea
&&a^2 = \ft{1}{12}\, (r^2 + \ell_1^2)\,,\qquad
b^2 = \ft{1}{12}\, (r^2 + \ell_2^2)\,,\qquad
c^2= \fft{r^2}{36h^2}\,,\nn\\
&& h^2=\fft{(r^2+ \ell_1^2)(r^2 + \ell_2^2)}{2r^4 + 3(\ell_1^2 +
\ell_2^2)\, r^2 + 6\ell_1^2\, \ell_2^2}\,,\qquad
d\rho=h\, dr\,.
\eea
This solution, a more general version of the metric with
$\ell_1=\ell_2$ obtained in \cite{b2,p2}, was constructed in
\cite{pando2} in a different coordinate system.  The radial coordinate
runs from 0 to $\infty$, with the geometry of $R^2\times S^2\times
S^2$ at small distance and the cone over $T^{1,1}/Z_2$
asymptotically. The metric describes a complex-line bundle over
$S^2\times S^2$.  Since the $\psi$ in (\ref{t11}) becomes the circular
coordinate of $R^2$ as $r\rightarrow 0$, it has a period of $2\pi$.
Thus, the principal orbit is a $T^{1,1}/Z_2$ instead of the $T^{1,1}$
space of the resolved conifold.  Although it may appear that the
metric reduces to a resolved conifold if one of the $\ell_i$ vanishes,
this is not the case since they have rather different principal
orbits.

      In this case there are, once again, three harmonic 2-forms.  As
in the previous case, we shall not consider the K\"ahler form.  The
one that carries non-trivial flux is given by (\ref{res2formans}) with
\be
u_1 = \fft{(\ell_1^2-\ell_2^2)(r^4-\ell_1^2\,\ell_2^2)}{(r^2 +
\ell_1^2)^2 (r^2+\ell_2^2)^2}\,,\quad
u_2 =\fft{r^4 + 2\ell_1^2\, r^2 + \ell_1^2\, \ell_2^2}{(r^2 +
\ell_1^2)^2 (r^2 + \ell_2^2)}\,,\quad
u_3 =\fft{r^4 + 2\ell_2^2\, r^2 + \ell_1^2\, \ell_2^2}{(r^2 +
\ell_1^2) (r^2 + \ell_2^2)^2}\,.\label{u123case3}
\ee
It is straightforward to verify that the 2-form is square
integrable at short distance but non-normalizable at large
distance.  The function $H$ is now given by
\vfil\eject
\bea
H\!\!\!&=&\!\!\! 1 + \fft{m^2\, (\ell_1^2-\ell_2^2)^2
(4r^2+\ell_1^2+\ell_2^2)}{4(\ell_1^2-3\ell_2^2)(3\ell_1^2-\ell_2^2)
(r^2 + \ell_1^2)(r^2 +\ell_2^2)}\\
\!\!\!&&\!\!\!+\fft{m^2\, (\ell_1^2+\ell_2^2)}{2\sqrt3
((\ell_1^2-3\ell_2^2)(3\ell_1^2-\ell_2^2))^{3/2}}\,
\log\Big(\fft{4r^2 + 3\ell_1^2 + 3\ell_2^2 -
\sqrt{3(\ell_1^2-3\ell_2^2)(3\ell_1^2-\ell_2^2)}}{
4r^2 + 3\ell_1^2 + 3\ell_2^2 +
\sqrt{3(\ell_1^2-3\ell_2^2)(3\ell_1^2-\ell_2^2)}}\Big)\,.\nn
\eea
When $\ell_1^2=3\ell_2^2$ or $\ell_2^2=3\ell_1^2$, the solution
becomes particularly simple because there is no longer a logarithmic
term.  Without the loss of generality, we set $\ell_2=\ell_1/\sqrt3$:
\be
H=1 + \fft{m^2\, (27r^6 + 63\ell_1^2\, r^4 + 45\ell_1^4\, r^2 +
11\ell_1^6)}{72(r^2+\ell_1^2)(3r^2 + \ell_1^2)}\,.
\ee
The solution for all non-vanshing $\ell_i$ is regular everywhere, with
$H$ as a positive constant at $r=0$ and behaving like (\ref{nutd3}) at
large $r$.  It is worth mentioning that, although the $u_i$ in
(\ref{u123case3}) reduce to the previous case when one of the $\ell_i$
vanishes, the same does not hold for the function $H$.  This is
because, in each case, the $H$ presented is not the most general
solution but rather has one of the integration constants chosen such
that the solution is regular.  This does not commute with setting
$\ell_i$ equal to zero.  As emphasized earlier, it does not come as a
surprise that the resolved conifold (over $T^{1,1}$) cannot be
obtained from the resolved cone on $T^{1,1}/Z_2$ in the limit of
vanishing $\ell_i$.

      The other harmonic 2-form has vanishing flux, given by
\be u_1 = \fft{(2r^2 + \ell_1^2 +\ell_2^2)}{(r^2 + \ell_1^2)^2
(r^2+\ell_2^2)^2}\,,\quad u_2 = \fft{1}{(r^2 +
\ell_1^2)^2(r^2+\ell_2^2)}\,,\quad u_3 = -\fft{1}{(r^2 +
\ell_1^2)(r^2 + \ell_2^2)^2}\,. \label{u123case4} \ee
In this case, we find that the 2-form is normalizable:
\be
\int_{0}^{\infty} \sqrt{g}\, L_\2^2 =
\fft{\ell_1^2 + \ell_2^2}{864\,\ell_1^4\,\ell_2^4}\,,
\ee
ensuring that the function $H$ is well-behaved at both large and small
$r$.  This function is given by
\bea H\!\!\!&=&\!\!\! 1 + \fft{\td m^2\,( (\ell_1^2-\ell_2^2)^2\,
r^2 + (\ell_1^2 + \ell_2^2)
((\ell_1^2-\ell_2^2)^2-\ell_1^2\,\ell_2^2))}{4\ell_1^4\, \ell_2^4\,
(\ell_1^2-3\ell_2^2)(3\ell_1^2-\ell_2^2)(r^2+\ell_1^2)(r^2+\ell_2^2)}
\\ \!\!\!&&\!\!\!  +\fft{\td m^2\,(\ell_1^2-\ell_2^2)}{2\sqrt3\,
\ell_1^4\,\ell_2^4\, ((\ell_1^2-3\ell_2^2)(3\ell_1^2-\ell_2^2))^{3/2}}
\,\log\Big(\fft{4r^2 + 3\ell_1^2 + 3\ell_2^2 -
\sqrt{3(\ell_1^2-3\ell_2^2)(3\ell_1^2-\ell_2^2)}}{ 4r^2 + 3\ell_1^2 +
3\ell_2^2 +
\sqrt{3(\ell_1^2-3\ell_2^2)(3\ell_1^2-\ell_2^2)}}\Big)\,.\nn
\eea
Again, when $\ell_2=\ell_1/\sqrt3$, the function $H$ becomes simple,
given by
\be
H=1+\fft{m^2\, (18r^4 + 36\ell_1^2\, r^2 + 19 \ell_1^4)}{
16\ell_1^6\, (r^2+\ell_1^2)^3(3r^2+\ell_1^2)}\,.
\ee
Thus, we see that $H$ for all non-vanishing $\ell_i$ is regular
everywhere; it is constant at small distance and behaves like
(\ref{nonutd3}) at large distance.  This solution is of particular
interest.  The metric interpolates a product space of $M_3$ and $U(1)$
bundle over $R^2\times S^2\times S^2$ at short distance to
AdS$_5\times T^{1,1}/Z_2$ at large distance. This implies that, for
the dual field theory, both conformal and Lorentz symmetries are
broken in general but are restored in the UV limit.

        The existence of a normalizable harmonic 2-form in this
manifold can be understood by the following.  In the resolved cone
over $T^{1,1}/Z_2$, the original singular apex is blown up to a smooth
$S^2\times S^2$, implying a non-collapsing 4-cycle.  Since there is no
supersymmetric 4-cycle in $T^{1,1}/Z_2$, it follows that there can
exist a normalizable harmonic 4-form.  In six dimensions, a 4-form is
Hodge dual to a 2-form, which we used to construct the wrapped
D3-brane.

        Both solutions are supersymmetric, since the functions $u_i$'s
in both cases satisfy the supersymmetric condition (\ref{susycon}).

       As we have shown that in the resolved conifold (and also in the
resolved cone over $T^{1,1}/Z_2$), there are two harmonic 2-forms.
Since the two harmonic 2-forms have the same structure
(\ref{res2formans}) and both satisfy the supersymmetric condition
(\ref{rescon}), they can be linearly superposed to give rise to a more
general $S^1$-wrapped D3-brane.  It is also possible that the D3-brane
wraps on a two-torus with the coordinates $x_2$ and $x_3$ fibred over
the two 2-forms respectively.

\section{On the deformed conifold}

       There is an alternative resolution to the singular conifold,
called the deformed conifold \cite{candel}.  The corresponding metric
is given by
\be
ds_6^2 = d\rho^2 + a^2\,\Big((\Sigma_1 + \sigma_1)^2 +
(\Sigma_2 + \sigma_2)^2\Big) + b^2\, \Big((\Sigma_1 - \sigma_1)^2
+(\Sigma_2 - \sigma_2)^2\Big) + c^2\, (\Sigma_3-\sigma_3)^2
\,,
\ee
where $a$, $b$ and $c$ are functions only of the radial variable
$\rho$. They satisfy the first-order equations
\be
\dot a =\fft{b^2 + c^2-a^2}{4b\,c}\,,\quad
\dot b =\fft{a^2 + c^2-b^2}{4a\,c}\,,\quad
\dot c =\fft{a^2 + b^2-c^2}{2a\,b}\,.
\ee
There is only one regular solution, given by
\bea
&&a^2 = \ft12 K\, \cosh^2(\ft12r)\,,\qquad
b^2 = \ft12 K\, \sinh^2(\ft12r)\,,\qquad
c^2 = \fft1{3K^2}\,,\nn\\
&& K=\fft{(\sinh(2r) - 2r)^{1/3}}{2^{1/3}\,\sinh r}\,,\qquad
h^2=\fft{1}{3K^2}\,,\qquad d\rho=h\, dr\,.
\eea
Now there exist only two harmonic 2-forms. One is the K\"ahler form
\be
J=e^0\wedge e^3 - e^1\wedge e^5 + e^2\wedge e^4\,,
\ee
where we define the vielbein basis
\bea
&&e^0=dt\,,\quad e^1=a\, (\Sigma_1+\sigma_1)\,,\quad
e^2=a\, (\Sigma_2+\sigma_2)\,,\quad
e^3=c\, (\Sigma_1-\sigma_1)\,,\nn\\
&&e^4=b\, (\Sigma_1-\sigma_1)\,,\quad
e^5=a\, (\Sigma_1-\sigma_1)\,.
\eea
The other harmonic 2-form is given by
\be
L_\2=\fft{1}{\sinh(2r) -2r}
(e^0\wedge e^3 + \ft12 e^1\wedge e^5 -\ft12 e^2\wedge
e^4)\,,\label{def2form}
\ee
which behaves like (\ref{2nd2form}) at large distance but is not
square integrable at small distance.  This is to be expected, since
the deformed conifold does not have any non-collapsing 2-cycles or
4-cycles.  The function $H$ behaves like (\ref{nonutd3}) at large
distance but the metric has a naked singularity at short distance.

     Note that, for (\ref{def2form}), the supersymmetric condition
(\ref{susycon}) is still satisfied and hence our solution is
supersymmetric, albeit the naked singularity.

     Since the deformed conifold has a non-collapsing 3-cycle, the
natural resolution is that of a fractional D3-brane, supported by a
self-dual harmonic 3-form.  Indeed, the fractional D3-brane on the
deformed conifold is regular and supersymmetric \cite{klebstra}.

\section{Conclusions}

      Since the conifold supports both harmonic complex self-dual
3-forms and 2-forms, there are two ways to add an additional flux
contribution to the D3-brane.  The first construction is to utilize
the 3-form, which gives rise to fractional D3-branes. In this paper,
we consider the second possibility, which utilizes the 2-form by
wrapping the D3-brane on $S^1$, which is fibred over the conifold.
The deformed conifold has a non-collapsing supersymmetric 3-cycle and,
consequently, the fractional D3-brane is regular, whereas the wrapped
D3-brane has a singularity at small distance.  On the other hand, the
resolved conifold has a non-collapsing, supersymmetric 2-cycle.  Thus,
the fractional D3-brane is singular, whilst the wrapped D3-brane is
regular. In both deformed and resolved conifolds, the 2-forms and
3-forms are not normalizable, due to the integrability at either large
or small distance.

       We also consider the resolved cone over $T^{1,1}/Z_2$.  In this
case, the apex singularity is blown up to a smooth $S^2\times S^2$.
Consequently, the manifold has a normalizable harmonic 4-form.  In the
six-dimensional transverse space, the 4-form is Hodge dual to a
2-form, which we use to construct a wrapped D3-brane.  The resulting
solution is regular everywhere, interpolating a product space of $M_3$
and $U(1)$ bundle over $R^2\times S^2\times S^2$ at short distance to
AdS$_5\times (T^{1,1}/Z_2)$ at large distance.  From the viewpoint of
the dual field theory, this implies that the broken conformal and
Lorentz symmetries are both restored in the UV limit.

       We argue that the seemingly different fractional D3-brane and
$S^1$-wrapped D3-brane can be united by T-duality as the same regular,
modified M2-brane on a Spin(7) manifold, which gives rise to the
resolved and deformed conifolds in different Gromov-Hausdorff limits.
Recent results for the $G_2$ unification of resolved and deformed
conifolds \cite{cglpunif1,brand,cglpunif2} strengthens the argument.
Thus, from the M-theory viewpoint, the fractional D3-branes are
natural for the deformed conifold, whilst the wrapped D3-branes
constructed in this paper are natural for the resolved conifold.


\begin{thebibliography}{99}

\bibitem{malda} J. Maldacena, {\sl The large $N$ limit of
superconformal field theories and supergravity},
Adv. Theor. Math. Phys. {\bf 2} (1998) 231, hep-th/9711200.

\bibitem{gkp} S.S. Gubser, I.R. Klebanov and A.M. Polyakov, {\sl Gauge
theory correlators from non-critical string theory}, Phys. Lett. {\bf
B428} (1998) 105, hep-th/9802109.

\bibitem{wit} E. Witten, {\sl Anti-de Sitter space and holography},
Adv. Theor. Math. Phys. {\bf 2} (1998) 253, hep-th/980215.

\bibitem{dlps} M.J. Duff, H. L\"u, C.N. Pope and E. Sezgin,
{\it Supermembranes with fewer supersymmetries,}
Phys. Lett. {\bf B371} (1996) 206, hep-th/9511162.

\bm{klebtsey} I.R. Klebanov and A.A. Tseytlin, {\sl Gravity duals of
supersymmetric $SU(N)\times SU(N+m)$ gauge theories},
Nucl. Phys. {\bf B578} (2000) 123, hep-th/0002159.

\bm{klebstra} I.R. Klebanov and M.J. Strassler, {\sl Supergravity and a
confining gauge theory: duality cascades and $\chi$SB-resolution of
naked singularities}, JHEP {\bf 0008} (2000) 052, hep-th/0007191.

\bm{ganpol} M. Gra\~na and J. Polchinski, {\it Supersymmetric
three-form flux perturbations on AdS$_5$}, Phys. Rev {\bf D63}
(2001) 026001, hep-th/0009211.

\bm{gubser} S. Gubser, {\it Supersymmetry and F-theory
realization of the deformed conifold with three-form flux},
hep-th/0010010.

\bibitem{cglpns2d2} M. Cveti\v c, G.W. Gibbons, H. L\"u and C.N. Pope,
{\it Supersymmetric non-singular fractional D2-branes and NS-NS
2-branes,} Nucl. Phys. {\bf B606} (2001) 18, hep-th/0101096.

\bm{candel} P. Candelas and X.C. de la Ossa, {\sl Comments on
conifolds},  Nucl. Phys. {\bf B342} (1990) 246.

\bm{zaytse} L.A. Pando-Zayas and A.A. Tseytlin, {\it 3-branes on a
resolved conifold}, JHEP {\bf 0011} (2000) 028,hep-th/0010088.

\bibitem{cglptrans}
M. Cveti\v c, H. L\"u and C.N. Pope,
{\it Brane resolution through transgression,}
Nucl. Phys. {\bf B600} (2001) 103, hep-th/0011023.

\bibitem{cglpsten} M. Cveti\v c, G.W. Gibbons, H. L\"u and C.N. Pope,
{\it Ricci-flat metrics, harmonic forms and brane resolutions,}
hep-th/0012011.

\bibitem{lvres} H. L\"u and J.F. V\'azquez-Poritz,
{\it Resolution of overlapping branes,} hep-th/0202075.

\bm{b2} L. Berard-Bergery, {\it Quelques examples de varietes
riemanniennes completes non-compactes a courbure de Ricci positive},
C.R. Acad. Sci. Ser. {\bf I302} (1986) 159.

\bm{p2} D.N. Page and C.N. Pope, {\it Inhomogeneous Einstein metrics
on complex line bundles}, Class. Quantum Grav. {\bf 4} (1987) 213.

\bm{pando2} L.A. Pando-Zayas and A.A. Tseytlin,
{\it 3-branes on spaces with $R \times S^2 \times S^3$ topology},
Phys. Rev. {\bf D63} (2001) 086006, hep-th/0101043.

\bm{deklm} M.J. Duff, J.M. Evans, R.R. Khuri, J.X. Lu and R. Minasian,
{\it The octonionic membrane}, Phys. Lett. {\bf B412} (1997) 281,
hep-th/9706124.

\bm{htr} S.W. Hawking and M.M. Taylor-Robinson, {\sl Bulk charges in
eleven dimensions}, Phys. Rev. {\bf D58} (1998) 025006,
hep-th/9711042.

\bm{becker} K. Becker, {\it A note on compactifications on spin(7)
manifolds}, JHEP {\bf 0105} (2001) 003, hep-th/0011114.

\bm{beckers} K. Becker and M. Becker, {\it M-theory on
eight-manifolds}, Nucl. Phys. {\bf B477} (1996) 155, hep-th/9605053.

\bibitem{hk} C.P. Herzog and I.R. Klebanov,
{\it Gravity duals of fractional branes in various dimensions,}
Phys. Rev. {\bf D63} (2001) 126005, hep-th/0101020.

\bibitem{cglphyper} M. Cveti\v c, G.W. Gibbons, H. L\"u and C.N. Pope,
{\it Hyper-K\"ahler Calabi metrics, $L^2$ harmonic forms, resolved
M2-branes, and AdS$_4$/CFT$_3$ correspondence,} Nucl. Phys. {\bf
B617} (2001) 151, hep-th/0102185.

\bibitem{cglpspin7} M. Cveti\v c, G.W. Gibbons, H. L\"u and C.N. Pope,
{\it New complete non-compact Spin(7) manifolds,} Nucl. Phys.{\bf
B620} (2002) 29, hep-th/0103155.

\bibitem{cgllp}
M. Cveti\v{c}, G.W. Gibbons, J.T. Liu, H. L\"u and C.N. Pope,
{\it A new fractional D2-brane, $G_2$ holonomy and T-duality,}
hep-th/0106162.

\bibitem{cglpunif1}
M. Cveti\v c, G.W. Gibbons, H. L\"u and C.N. Pope,
{\it M-theory conifolds}, hep-th/0112098, to appear in
Phys. Rev. Lett.

\bibitem{brand} A. Brandhuber,
{\it $G_2$ holonomy spaces from invariant three-forms},
hep-th/0112113.

\bibitem{cglpunif2}
M. Cveti\v c, G.W. Gibbons, H. L\"u and C.N. Pope,
{\it A $G_2$ unification of the deformed and resolved conifolds},
hep-th/0112138.


\end{thebibliography}
\end{document}